# A practicable method for the analysis of complex motion of biological and soft matter

Jun Ma[*]
*School of Science, Hubei University of Technology, Wuhan 430068, P. R. China*

Biological function of living matter is fulfilled by complex motions of biological and soft matter. Unlike general motion is deterministic described by Newton's laws, these motions are mostly random and uncertain for the position in stochastic process, being characterized as irregular trajectories of movement without a defined velocity. Like human fingerprint, the trajectory is the identity of the motion containing fundamental dynamical information. Such irregular trajectories randomly interwind and twist to each other to produce a complicated turmoil configuration in which so far the unrealized mechanism of motion is hidden. Nowadays, the analytical method for this 'fingerprint' trajectory is still missed. Here we develop a practicable method to decipher complicated trajectory configuration, which uncovers abundant dynamical information hiding in irregular trajectories, revealing the remarkable evolution of spatio-temporal microstructure, thus leading to the novel systematic study of the dynamics of biological and soft matter.

The functions of life in biological body are fulfilled by a variety of the dynamical motions of biological matter. The character of this dynamics of biological matter is the stochastic process which is dominated by the random motion such as: SARS-CoV-2 virus exploring its way to attach to human cell [1]; protein folding [2]; cell migration [3]; ligand binding receptor [4]; microorganism (bacteria) transport [5]; enzyme binding the substrate [6]; internal motion of supercoiled DNA [7]; protein searching for DNA target sites [4]; molecular motor (protein Kinesin-1) moving on microtubules [8]; vesicular cargo delivering to a subcellular domain [4]; etc.

Random motions, typically Brownian motion, displays irregular trajectories of movement without a defined velocity due to the non-differentiability of zigzag trajectories (Fig.1a). Its dynamics is described by Einstein that the mean-squared displacement of a particle grows linearly with time [9]. However, this ensemble statistics does not give the details of the affluent diversity of individual movements in the dynamical process.

Looking quite mess and disorder, the zigzag trajectories interwind to each other to constitute a very complex configuration, which hides the inherent cryptogram-like mechanism. The irregular trajectories contain vast intrinsic dynamical information in details being the fundamental characters of random motions, like the fingerprint representing the unique identity of human. Nowadays, the analysis method for these "fingerprint-identity" trajectories is still missed. In this work, we develop a practicable method which can decipher the dynamical information of irregular trajectories.

Irregular trajectories indicate that the position of particle in Brownian motion can not been determined exactly, which leads to the probability distribution to describe a spatial range of its position. This concept of uncertainty and possibility description is similar to that of quantum mechanics, thus Nelson [11] and Comisar [12] derived a kinetic equation from Brownian motion and showed its being equivalent to Schrödinger equation, indicating an intrinsic connection between stochastic process and quantum mechanics. However, those kinds of theories were not been developed to study biological matter which is soft condensed matter. The real difficulty of this route is although the dynamical behaviour of soft matter is rarely deterministic, but being random, the time scale of quantum fluctuations is much smaller than the response time of soft matter, thus the quantum effect of soft matter is negligible [13].

Facing this challenge, in this work, we develop a method to quantize the stochastic motion of the biological and soft matter. Through the quantization, it reveals the vast dynamical fingerprint-like information stored in complicated turmoil trajectories in stochastic process. On the other hand, the significant character of microscopic world is the quantum fluctuation marked by the trajectories of random motions, e.g., the electron performs Brownian motion around an atom with very irregular trajectories [11]. Although "static" electron cloud is actually produced by these dynamical trajectories which are the most straightway first-hand information indeed, those trajectories can not be observed in the quantum experiment, thus deciphering the information of irregular trajectories may shed light on the future theoretical development on quantum mechanics.

We discuss one dimensional (1D) case of spherical Brownian particle in fluid (3D case can be decomposed on *x, y, z* coordinate axis). For the time scale $t_b < m/6\pi\eta R$ (*m*, particle mass; *R*, particle radius; $\eta$, the viscosity of fluid), any segment of trajectory consists of a number of steps (Fig. 1a. inset) produced by the ballistic motion [14] with the mean velocity $\langle v_0 \rangle = \sqrt{\frac{8kT}{\pi m^*}}$ [15] ($m^*$ is an effective mass being the mass of particle plus half the mass of the displaced fluid [14]). Einstein gave the time for one step of ballistic motion as [16]: $t_b = m^* \ln 10/6\pi k_B R$ ($k_B$, Boltzmann constant), thus the mean step length of ballistic motion is $l = \langle v_0 \rangle t_b$.

The Brownian particle is considered to be bashed in the random field produced by the random thermal motion of the fluid molecular, thus the chance to move along the positive or negative direction is the same. Li experimentally reported that the Brownian particle oscillated around its equilibrium position resulting from it performing the ballistic motion with flipping the motion direction (Fig.1b) [10]. Actually, due to the fluctuation of the random field, in a time interval $\tau$, *N* ballistic steps include extra *M* steps along positive (or negative) direction and equivalent $(N-M)/2$ steps along the positive and negative direction respectively [17]. In each step, the velocity initially increases to *v*, then gradually decays. A positive step and a negative step produce no net displacement, which is equivalent to one-periodical oscillation. Therefore, in the time interval $\tau$, for the ensemble average of the whole motion, a couple of $(N-M)/2$ positive and negative steps are equivalent to the oscillations in $(N-M)/2$ periods. This equivalence of the ensemble average is schematically shown in Fig.1c.

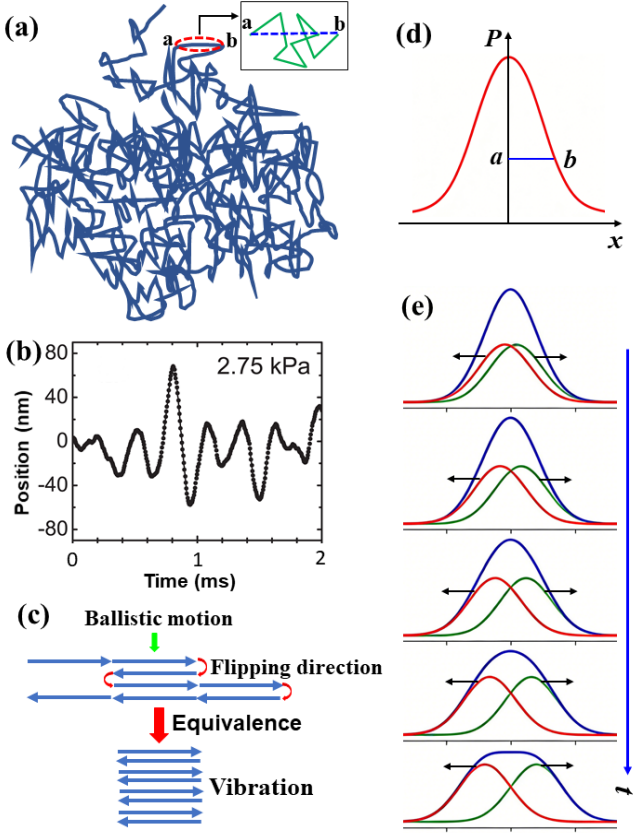

Figure 1: (a) Irregular trajectories of Brownian motion, inset: a segment of trajectory 'ab' consisted of a number of steps produced by the ballistic motion of Brownian particle. (b) One-dimensional trajectories of a 3 $\mu m$ diameter silica bead in air at 2.75 kPa performs ballistic motion around the equilibrium position '0' due to consistently flipping the motion direction [10]. (c) The ensemble effect of random ballistic motion is equivalent to a vibration. An arrow represents one step of ballistic motion. (d) The segment of trajectory (e.g.,'ab') is the displacement ($x$) of particle observed under optical microscopy, whose probability density $P$ displays a Gaussian distribution. (e) The fluctuation field produces two wave packets propagating along the right and left directions (shown by the arrows) respectively, the blue curve is the superimposition of those two.

For each period, considering a system of a Brownian particle immersed in the fluid, we can establish the effective potential of particle in the thermal bath: initially the velocity of particle increases resulting form the thermal motion of fluid molecules imparting the energy to the particle by the collision, which can be considered as thermal bath (potential) acting work on the particle; then the velocity of particle gradually decreases due to particle transferring its energy to fluid molecules by the collision, which corresponding to the kinetic energy of particle converting to the energy (potential) of thermal bath. With above construction, we approximately consider the process as quasi-harmonic oscillation.

In the time interval $\tau$, the motion of Brownian particle is the vibration for $(N-M)/2$ periods plus the net displacement $Ml$. Since the vibration direction is the same as the net displacement direction, for the ensemble average during $\tau$, this is a approximate longitudinal wave moving along $x$ axis. On the Gaussian distribution of particle displacement shown in Fig.1d, each $x$ corresponds to a propagation displacement ($M$ steps) of the wave in $(N-M)/2$ periods, it has the period $T_p = 2\tau/(N-M)$, wavelength $\lambda = 2x/(N-M)$, the step number $N = \tau/t_b$.

In the time interval $\tau$, $M$ extra steps produce a net displacement $x$ (= $Ml$) which is a rectilinear segment in the trajectory, e.g. 'ab' in Fig.1a, which is 'one pace' of particle motion displayed under optical microscopy. Those rectilinear segments produced by $N$ Brownian particles in the time interval $\tau$ show a Gaussian function for the probability distribution $P$ (Fig.1d) [18]:

$$P(x,t) = \frac{1}{\sqrt{4\pi D\tau^\alpha}} \exp\left(-\frac{x^2}{4D\tau^\alpha}\right) = \frac{1}{\sqrt{2\pi}\sigma_0} \exp\left(-\frac{x^2}{2\sigma_0^2}\right) \tag{1}$$

where $D$ is the diffusion constant of particle, the standard derivation of Gaussian distribution $\sigma_0 = \sqrt{2D\tau^\alpha}$, $\alpha = 1$ for diffusion, $\alpha < 1$ subdiffusion, $\alpha > 1$ superdiffusion.

The probability distribution $P$ for Brownian particle is related to a wave function $\psi$ [19–21]: $P = \psi\psi^*$, $\psi^*$ is the conjugate function. Constructing the wave function at $t = 0$ by Equation (1), $\psi(x,0) = \frac{1}{\sqrt[4]{2\pi\sigma_0^2}}e^{-x^2/4\sigma_0^2+ik_0x}$, $k_0$ is the mean wave number of wave packet. It has:

$$\psi(x,t) = \int_{-\infty}^{+\infty} a_n(k)e^{\pm i(kx-\omega t)}dk \tag{2}$$

where $\omega$ is angular frequency, $k$ wave number. we have $a_n(k) = \frac{\sqrt{\sigma}}{\sqrt[4]{2\pi^3}}\exp(-\sigma^2(k-k_0)^2)$ and $\psi$ (see Appendices),

$$\psi(x,t) = A\exp\left(-\frac{(x-v_gt)^2}{4\sigma_0^2+(\beta t/\sigma_0)^2}\right)\exp(\pm i(k_0x-(\omega_0-\phi)t)) \tag{3}$$

where $v_g = \frac{d\omega}{dk}|_{k_0} = \frac{\pi^2 nl}{(k_0 l+\pi)^2}$ is group velocity ($n = N/\tau$ is the number of steps per unit time), $\omega_0$ the mean angular frequency of wave packet, $\beta = \frac{d^2\omega}{dk^2}|_{k_0} = -\frac{2n(\pi l)^2}{(k_0l+\pi)^3}$, $A = \frac{\sqrt[4]{2/\pi}}{\sqrt{2\sigma_0 \pm i\beta t/\sigma_0}}$, $\phi = \frac{\beta(x-v_gt)^2}{2(4\sigma_0^4+(\beta t)^2)}$, $k_0 = \frac{\pi(\sqrt{2\pi}Nl-\sigma_0)}{\sigma_0 l}$. For '$\pm$' in Equation (2) and (3), '+' and '−' corresponds to the wave functions $\psi_+$ and $\psi_-$ moving along the positive and negative directions, respectively.

From $P = \psi\psi^*$, we get the probability distribution of particle,

$$P(x,t) = \frac{1}{\sqrt{\pi(2\sigma_0^2+(\beta t/\sqrt{2}\sigma_0)^2)}}\exp\left(-\frac{(x-v_gt)^2}{2\sigma_0^2+(\beta t/\sqrt{2}\sigma_0)^2}\right) \tag{4}$$

The displacement of wave packet motion is the displacement of particle in diffusion. Since the wave packet moves with the group velocity, and the center of wave packet is the mean position of wave components making up wave packet, i.e., the mean position of Brownian particle, thus the mean displacement of Brownian particle is the displacement of the center of wave packet $<x>=v_g t$, it has (see Appendices),

$$\langle x \rangle = \frac{\pi^2 nlt}{(k_0 l + \pi)^2} \quad (5)$$

Equation (5) indicates that the mean displacement of particle (or the center of wave packet) increases with decreasing the mean wave number $k_0$ and increasing the step numbers in unit time $n$ and the mean step length $l$. Thus the movement of particle in diffusion is closely related to the ballistic motion by $n$, $l$ and the carrier wave of wave packet by $k_0$.

Considering $j$ displacements of Brownian particles with Gaussian distribution (Fig.1d), by using Equations (1)(2), we can construct the corresponding $j$ wave components constituting a wave packet with $j$ displacements (propagation distances), thus Fig.1d includes the information of spatial frequency spectrum of wave packet. The Gaussian probability distribution of displacement shows the uncertainty of particle displacement, this "uncertainty" is converted into the width of spatial frequency $\Delta k$ of wave packet. Thus the wave packet is adopted to represent the particle performing random motion, which is similar to the position uncertainty of quantum particle described by a wave packet, e.g., the motion of free electron is represented by wave packet propagating [22].

From Equation (3), we can transform the movement of Brownian particle with Gaussian probability distribution into the propagation of wave packet represented by a wave function. The phase of wave function corresponds to the mean wave number $k_0$ and the angular frequency $\omega_0 - \phi$ of wave packet. As indicated by Equation (3), with $t$ increasing, the wave packet disperses during the propagation process, which corresponds to the position range of particle distribution enlarging as diffusion proceeding. From Equation (4), the position distribution of particle is Gaussian distribution with the mean position $v_g t$ and the standard derivation describing the position range of the particle distribution; the dispersion of wave packet leads to the standard derivation $\sigma$ increasing with $t$ increasing,

$$\sigma = \sqrt{\sigma_0^2 + (\beta t / 2\sigma_0)^2} \quad (6)$$

For the view of the field, Brownian motion results from a random field which stems from the random collision of the fluid molecules, thus the chance for the wave packet (particle) moving along the positive or negative direction is equal (Fig.1e). However, the net displacement of wave packet (particle) is produced by the fluctuation of random field. The Gaussian diffusion starting from a position can be divided into two Gaussian wave packets simultaneously moving along the positive and negative directions respectively (Fig.1e).

Here we discuss the motion along the positive direction. If the wave packet described by Equation (3) is moving along the positive direction all the time, it should reach to the position $A'$ in Fig.2a, but actually it arrives at the position $A$. The reason is that the motion of wave packet flips between the positive and negative directions corresponding to the wave function $\psi_+$ and $\psi_-$ respectively, but with more paces along the positive direction produced by the fluctuation of random field along the same direction. As shown in Fig.2b, the wave packet oscillates forward (pace ① or ③) and backward (pace ② or ④) with one more positive pace (⑥) producing a net positive displacement $d$. With this net displacement accumulating as the time increasing, it leads to the displacement of wave packet motion (corresponding to particle diffusion) $L$.

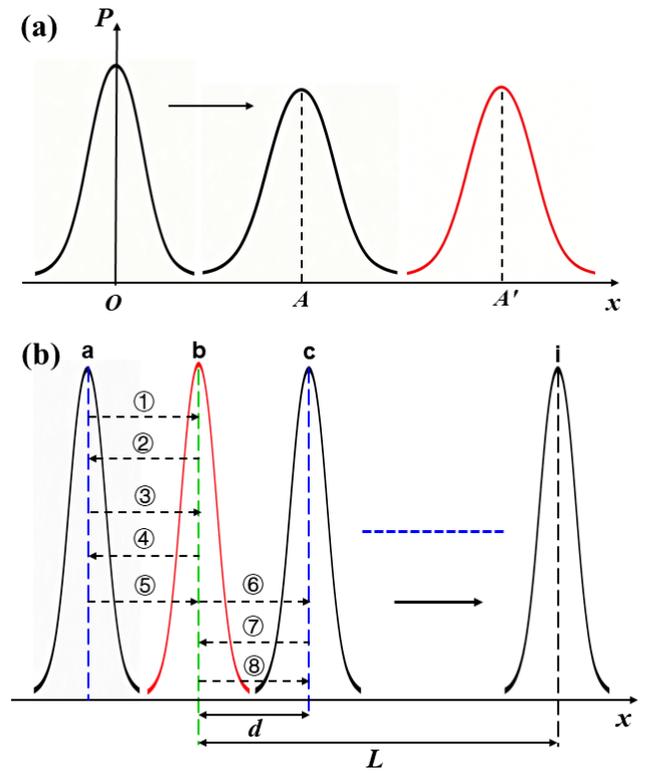

Figure 2: (a) The wave packet moving along $x$ axis is expected to reach the position $A'$, but it actually arrives at the position $A$ due to the oscillating motion. (b) The propagation mechanism of wave packet: the oscillating motion plus drift motion. A oscillating motion consists of the forward motion ① and the backward motion ②, so as to the motion ③④. After the oscillation motion ①②③④⑤, the wave packet drifts to ⑥ due to a fluctuation of random field along the positive $x$ direction, producing a drift motion with a displacement $d$. The wave packet is continually repeating above process until producing the net displacement $L$, which corresponds to the displacement in (a) with $OA = L$

This oscillation results in the displacement of wave packet shrinking from the position $OA'$ to $OA$ (Fig.2a). If the time of Brownian motion is $t_T$, $t_T = t_o + t$, $t_o$ and $t$ are the time corresponding to the oscillation and the net displacement respectively, $t_o = AA'/v_g$, $t = OA/v_g$, $OA' = v_g t_T$.

For any given time $t_T$, excluding the time $t_o$ for wave packet oscillating forward and backward, the actual time for wave packet to produce the net displacement is $t$ being used in Equations (2)(3)(4)(5)(6). Since the wave packet performs the random motion with the net displacement, it has (see Appendices).

$$t = \sqrt{t_T} \quad (7)$$

From Equation (5), we know that the mean displacement of particle is unrelated to the mass of particle (see Appendices), which is consistent with Ref. [9]. The parameter $t$ reflects the extent and scale of the net displacement, from which the position of wave packet (or the mean position of Brownian particle) can be obtained. There is no dispersion for the wave packet performing the oscillation motion [23]. Here the dispersion of wave packet along the positive motion cancels with that along the negative motion, but the dispersion is accumulated for the net positive motion given by $\sigma$ in Equation (6) corresponding to the time $t$.

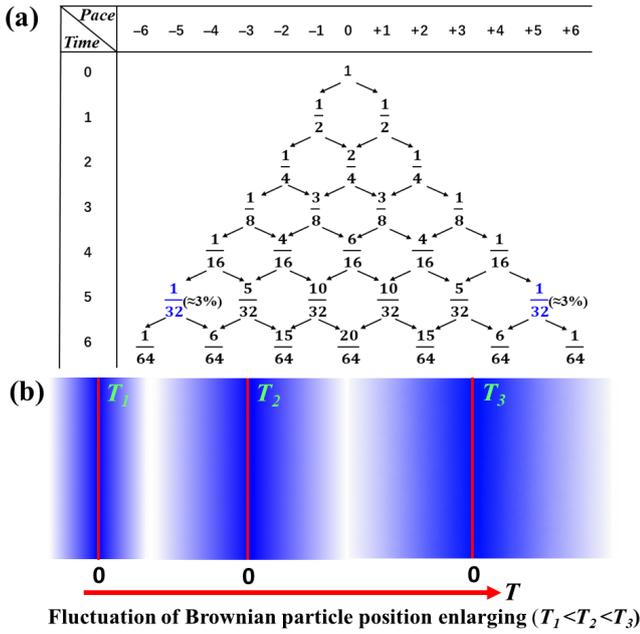

Figure 3: The fluctuation of wave packet (particle) position due to the asymmetry of the forward and backward motion of wave packet. (a) Symmetrical distribution of the forward and backward motion of wave packet, represented by the pace numbers. '0' pace is the mean position of wave packet (particle) with no fluctuation. The probability of wave packet (particle) distribution decreases with the pace number increasing. (b) The position fluctuation of wave packet (particle) enlarges with temperature increasing. '0' corresponding to '0' pace in (a). With the pace number increasing at the two sides of '0', the probability of particle distribution decreases, representing with blue color becoming lighter at the two sides of '0'.

The above description for wave packet in Fig.2b corresponds to the observation under optical microscopy, the direction of particle motion randomly switches between the positive and negative directions but with more positive paces leading to the diffusion along the positive direction. The only difference is that the pace length of wave packet is the mean displacement of particle, i.e., the mean pace length under optical microscopy, which derives from constructing the wave packet from the $N$ particle displacements with Gaussian distribution (Fig.1d). We denote the displacement of the center of wave packet in time $\tau$ as a pace, the length of a pace is (see Appendices),

$$\sigma_0 = \sqrt{2D\tau^\alpha} = \sqrt{\frac{k_B T \tau^\alpha}{3\pi\eta R}} \qquad (8)$$

Since the wave packet consisting of $j$ component waves corresponds to $j$ Brownian particles, and the center of wave packet is the mean position of $j$ component waves (or particles), thus $\sigma_0$ is the mean length of the pace for particle motion observed under the optical microscopy. As indicated by Equation (8), $\sigma_0$ increases with the increase of temperature, and decreases with the increase of particle radius.

For the ideal condition of Brownian particle under the random field, the chance to move to the positive and negative direction is equal, thus the particle consistently flips the direction of movement. Here the particle is represented by the wave packet in Fig.3a: the wave packet oscillates forward and backward between the position '+1' and '-1', corresponding to a positive pace and a negative pace from the equilibrium position '0', respectively.

But actually the random field fluctuates from time to time, producing more positive paces than negative ones, vice versa, thus the wave packet could consecutively move the positive or negative paces, which leads to the fluctuation of the position of wave packet.

As indicated in Fig.3a, the probability of the successive positive or negative paces deceases with the consecutive number of paces increasing: for the movement of 5 paces, the probability of wave packet moving to the position of +1, +3, +5 (or -1, -2, -3) paces correspondingly decreases to 5/16, 5/32, 1/32, respectively, which is represented by blue color gradually becoming lighter at the two sides of the position '0' in Fig.3b. The probability is 94% for the position fluctuation of wave packet ranging from -5 paces ($-5\sigma_0$) to +5 paces ($5\sigma_0$), so we approximately think that the wave packet (particle) locates in the range from $-5\sigma_0$ to $5\sigma_0$.

Due to the length of pace $\sigma_0 \propto \sqrt{T}$, the position fluctuation (represented by the blue color in Fig.3b) of wave packet enlarges with temperature increasing. Since the wave packet corresponds to the probability distribution of Brownian particle, we can know the position fluctuation of the particle from the position fluctuation of the wave packet. The fluctuation distance of the particle is rather than the continual value but a "quantum" (integer) number multiplying with $\sigma_0$.

The merit of the method is as follows. It can extract the dynamical information contained in trajectories on two spatio-temporal scales. On the smaller spatio-temporal scale for ballistic motion, the dynamical information is included in the parameters $v_g, n, l, k_0$.

On the larger spatio-temporal scale of the particle diffusion, the Gaussian distribution of random displacements (the trajectory segments) is converted to the waves of different frequencies constituting the wave packet. The motion of wave packet flips between the positive and negative directions with the net motion along one direction, which is described by the parameters $t_o, t, \sqrt{t_T}$. These parameters reveals how frequently the particle rolls backward as it is proceeding forward. The distribution range of the particle is described by the standard derivation $\sigma$ given by Equation (6), whereas the fluctuation of particle position is a "quantum" (integer) number multiplying with $\sigma_0$ in Equation (8).

In summary, in this work we develop a practicable method which fully utilizes the dynamical information hiding in the complex trajectories of random motion. Comparing with the traditional macroscopic diffusion theory, the present theory is the microscopic one including the dynamical details for ballistic motion and particle trajectories, from which uncovers the fundamental evolution of spatio-temporal microstructure for Brownian particle: the mobility, the probability distribution, the dispersion of position distribution, the fluctuation of position distribution, etc.

## Appendices

1. Wave function

The displacements produced by $N$ Brownian particles in time $\tau$ show a Gaussian distribution $P$ [18]:

$$P(x,t) = \frac{1}{\sqrt{4\pi D \tau^\alpha}} \exp\left(-\frac{x^2}{4D\tau^\alpha}\right) = \frac{1}{\sqrt{2\pi}\sigma_0} \exp\left(-\frac{x^2}{2\sigma_0^2}\right) \quad (9)$$

where $D$ is the diffusion constant of particle, the standard derivation of Gaussian distribution $\sigma_0 = \sqrt{2D\tau^\alpha}$, $\alpha = 1$ for diffusion, $\alpha < 1$ subdiffusion, $\alpha > 1$ superdiffusion.

Constructing the wave function $\psi$ at $t = 0$ by Equation (9) with the relation $P = \psi\psi^*$, $\psi(x,0) = \frac{1}{\sqrt[4]{2\pi\sigma_0^2}} e^{-x^2/4\sigma_0^2 + ik_0 x} = \int_{-\infty}^{+\infty} a_n(k) e^{ikx} dx$, $k_0$ and $k$ are the mean wave number and wave number, respectively. Thus, $a_n(k) = \frac{1}{2\pi}\int_{-\infty}^{+\infty} \psi(x,0) e^{-ikx} dx = \frac{1}{2\pi}\int_{-\infty}^{+\infty} e^{-x^2/4\sigma_0^2 + ik_0 x} e^{-ikx} dx$, after calculate the integral, we get,

$$a_n(k) = \frac{\sqrt{\sigma_0}}{\sqrt[4]{2\pi^3}} \exp(-\sigma_0^2 (k-k_0)^2) \quad (10)$$

The wave function is,

$$\psi(x,t) = \int_{-\infty}^{+\infty} a_n(k) e^{\pm i(kx - \omega t)} dk \quad (11)$$

For '±' in Equation (11), '+' and '−' correspond to the wave functions $\psi_+$ and $\psi_-$ moving along the positive and negative directions, respectively.

For a Brownian particle having a net displacement along the positive $x$ axis in a time interval $\tau$, the total $N$ steps of ballistic motion is divided into $(N-M)/2$ steps along the positive and negative directions respectively and $M$ extra steps along the positive direction [17]. For the ensemble effect, this motion is equivalent to a wave propagating to a distance $\langle x_\tau \rangle = Ml$ in $(N-M)/2$ periods, $l$ is the mean step length of ballistic motion. Thus we have the period $T_p$, angular frequency $\omega$ and mean displacement (propagation distance) $\langle x_\tau \rangle$ as follows,

$$N = \tau/t_b \quad (12)$$

where $t_b$ is the time for one step of ballistic motion.

$$T_p = 2\tau/(N-M) \quad (13)$$

$$\omega = 2\pi/T_p = \pi(N-M)/\tau \quad (14)$$

$$\langle x_\tau \rangle = Ml \quad (15)$$

$$\lambda = \frac{2x_\tau}{N-M} = \frac{2Ml}{N-M} \quad (16)$$

$$k = \frac{\pi(N-M)}{Ml} \quad (17)$$

So the extra numbers of steps $M = \langle x_\tau \rangle / l$.

From Equation (17), we get $M = \frac{N\pi}{\pi + kl}$, substitute it into Equation (14), we get,

$$\omega = \frac{N\pi kl}{(kl+\pi)\tau} = \frac{\pi n kl}{(kl+\pi)} \quad (18)$$

where $n = N/\tau$ is the numbers of steps per unit time.

The angular frequency $\omega$ expands Taylor series and takes the first three terms:

$$\omega = \omega(k_0) + \frac{\partial \omega}{\partial k}|_{k_0}(k-k_0) + \frac{1}{2}\frac{\partial^2 \omega}{\partial k^2}|_{k_0}(k-k_0)^2 + ... \quad (19)$$

Take the derivative of Equation (18), we get,

$$\frac{d\omega}{dk}|_{k_0} = \frac{\pi^2 nl}{(k_0 l + \pi)^2} \quad (20)$$

$$\frac{d^2\omega}{dk^2}|_{k_0} = -\frac{2n(\pi l)^2}{(k_0 l + \pi)^3} \quad (21)$$

Substitute Equations (10)(19) into Equation (11), after the calculation, we have,

$$\psi(x,t) = A \exp\left(-\frac{(x-v_g t)^2}{4\sigma_0^2 + (\beta t/\sigma_0)^2}\right) \exp(\pm i(k_0 x - (\omega_0 - \phi)t)) \quad (22)$$

where $v_g$ is the group velocity, $\beta = \frac{d^2\omega}{dk^2}|_{k_0} = -\frac{2N(\pi l)^2}{\tau(k_0 l + \pi)^3}$, $A = \frac{\sqrt[4]{2/\pi}}{\sqrt{2\sigma_0 \pm i\beta t/\sigma_0}}$, $\phi = \frac{\beta(x-v_g t)^2}{2(4\sigma_0^4 + (\beta t)^2)}$, $k_0 = \frac{\sqrt{2\pi^3}(N-M)}{\sigma_0}$. For '±' in Equation (22), '+' and '−' corresponding to the wave functions $\psi_+$ and $\psi_-$ moving along the positive and negative directions, respectively.

2. The mean wavenumber of wave packet

Consider the wave packet moving along the positive direction, for the Gaussian distribution of displacements corresponding to the time interval $\tau$, the mean displacement

$$\langle x_\tau \rangle = \int_0^{+\infty} \psi x \psi^* dx = \int_0^{+\infty} \frac{x}{\sqrt{2\pi}\sigma_0} \exp\left(-\frac{x^2}{2\sigma_0^2}\right) dx = \frac{\sigma_0}{\sqrt{2\pi}} \quad (23)$$

$\langle x_\tau \rangle$ is the propagation distance of a wave with the mean wavenumber $k_0$ of wave packet.

The group velocity $v_g = \langle x_\tau \rangle / \tau = \omega/k_0$. Thus

$$k_0 = \omega \tau / \langle x_\tau \rangle \quad (24)$$

Substitute Equations (14)(15)(23) into Equation (24), we get the mean wavenumber of wave packet,

$$k_0 = \frac{\pi(\sqrt{2\pi} Nl - \sigma_0)}{\sigma_0 l} \quad (25)$$

3. The step length of ballistic motion

The ballistic motion with the mean velocity is [15],

$$\langle v_0 \rangle = \sqrt{\frac{8kT}{\pi m^*}} \quad (26)$$

$m^*$ is an effective mass being the mass of particle plus half the mass of the displaced fluid [14].

The time for one step of ballistic motion is given by [16]:

$$t_b = \frac{m^* \ln 10}{6\pi k_B R} \quad (27)$$

$k_B$, Boltzmann constant; $R$, the radius of particle.

The mean step length of ballistic motion is,

$$l = \langle v_0 \rangle \cdot t_b = \frac{\ln 10}{3\pi R}\sqrt{\frac{2Tm^*}{\pi k_B}} \quad (28)$$

4. The pace length of wave packet

The diffusion constant can be represented as [17],

$$D = \frac{k_B T}{6\pi \eta R} \quad (29)$$

Substitute Equations (29) into $\langle \sigma_0^2 \rangle = \sigma_0^2 = 2D\tau^\alpha$, we have the pace length of wave packet,

$$\sigma_0 = \sqrt{2D\tau^\alpha} = \sqrt{\frac{k_B T \tau^\alpha}{3\pi \eta R}} \quad (30)$$

where $\eta$ is the viscosity of fluid.

5. The mean displacement of Brownian particle

From Equation (20), we have the group velocity,

$$v_g = \frac{d\omega}{dk}|_{k_0} = \frac{\pi^2 nl}{(k_0 l + \pi)^2} \quad (31)$$

The mean displacement of Brownian particle is the displacement of the center of wave packet, that is,

$$\langle x \rangle = v_g t = \frac{\pi^2 nlt}{(k_0 l + \pi)^2} = \frac{\pi^2 Nlt}{\tau(k_0 l + \pi)^2} \quad (32)$$

For the $N$ paces of random motion with $(N-M)/2$ paces along the positive and negative directions respectively and $M$ extra paces along the positive or negative direction, the mean square displacement represented by the number of paces is [17], $\langle M^2 \rangle = N$. $M$ paces satisfy the Gaussian distribution, $\langle M^2 \rangle = \langle M \rangle^2 + \chi^2$, $\chi$ is the standard derivation of Gaussian distribution. Since $\langle M \rangle = 0$, thus has,

$$\langle M^2 \rangle = \chi^2 = N \quad (33)$$

If $N$ paces are completed in a time interval $t_T$, from Equation (33) the time for completing $\chi$ paces is $t = \sqrt{t_T}$. Since $\chi$ is the mean positive (or negative) paces corresponding to the mean displacement $\langle x \rangle (= \chi l)$ along the positive (or negative) direction, the time for performing the mean displacement $\langle x \rangle$ is,

$$t = \sqrt{t_T} \quad (34)$$

The time interval $\tau$ and $t_T$ can be represented by $t_b$ as: $\tau = pt_b$, $t_T = qt_b$, $p$, $q$ are the integers. Substitute $\tau$, $t_T$ and Equations (27)(28)(34) into Equation (32), we get,

$$\langle x \rangle = \frac{2\pi N}{p(k_0 l + \pi)^2}\sqrt{\frac{qT \ln 10}{3R}} \quad (35)$$

Equation (35) indicates that the mean positive (or negative) displacement of particle is unrelated to the mass of particle, which is consistent with Ref. [9].

The author appreciates Science Foundation of Hubei University of Technology (Grant NO.BSQD2017064) for supporting this work.